\documentclass[reprint,superscriptaddress,a4paper,aps,showpacs,longbibliography]{revtex4-1}
\usepackage{graphicx}
\graphicspath{{./},{/home/user/fig/publi/muon/}}
\usepackage{amsfonts}
\usepackage{amsbsy}
\usepackage{amsmath}
\usepackage{diagbox}

\begin{document}

\title{Why paramagnetic chiral correlations in the long wavelength limit do not contribute to muon-spin relaxation}

\author{A. Yaouanc}
\affiliation{Universit\'e Grenoble Alpes, CEA, IRIG-PHELIQS, F-38000 Grenoble, France}
\author{P. Dalmas de R\'eotier}
\affiliation{Universit\'e Grenoble Alpes, CEA, IRIG-PHELIQS, F-38000 Grenoble, France}

\date{\today}

\begin{abstract}
  A crystal structure that cannot be superposed on its mirror image by any combination of rotations and translations is classified as chiral. Such crystal structures have gained importance in recent years since they are prone to host unconventional magnetic orders and to exhibit topological magnetic textures. These properties result from the Dzyaloshinskii-Moriya antisymmetric exchange interaction which is authorized when space inversion is broken. While recent reports have shown the muon spin rotation and relaxation technique to provide unique information about structural and dynamical properties which are specific to chiral magnets in their ordered phase, the question here is whether this technique is sensitive to paramagnetic chiral correlations that are observed in neutron scattering experiments above the critical temperature. In the relevant long wavelength limit, it is shown that they do not contribute to the relaxation rate, which in turn only probes non-chiral correlations.

\end{abstract}

\maketitle

\section{Introduction} 
\label{Introduction}

Chirality (or handedness) is ubiquitous in nature. It occurs at the microscopic level and its effect was recognized in molecules by Pasteur at a time as early as the mid 19th century \cite{Pasteur48,Flack09}. A century later the break of symmetry between right and left was revealed at an even lower scale in the parity violation in weak interactions \cite{Lee56}. The muon spin rotation or relaxation ($\mu$SR) technique relies on it  \cite{Garwin57}. In magnetism, chirality has been reported, e.g., for magnetic structures \cite{Ishida85}, for solitons \cite{Braun05}, and geometrically frustrated compounds \cite{Plakhty00}.  A great deal of attention has been devoted to systems crystallising in the so-called non-centrosymmetric B20 structure and containing a $3d$ transition element and silicon or germanium \cite{Williams66}. The chiral link between structure and magnetism for these metals has been investigated \cite{Tanaka85,Ishida85,Dyadkin11,Dmitriev12}. More recently the non-centrosymmetric nature of a structure has been recognized to give birth not only to helical magnetic structures \cite{Ishikawa76}, but also to skyrmion lattices \cite{Muhlbauer09a} and isolated skyrmions \cite{Fert13}. Moreover these phenomena are not only found among metals and alloys \cite{Ishikawa76,Lebech89,Grigoriev10}, but also in semiconductors  \cite{Adams12} and multiferroics \cite{Lottermoser04,Goto04}, and devices based on magnetic chirality have been discussed \cite{Dor13}.

The relativistic Dzyaloshinskii-Moriya (DM) antisymmetric exchange interaction, which results from non-centrosymmetric crystal structures \cite{Dzyaloshinskii58,Moriya60}, is the culprit of chiral magnetic structures, excitations, and correlations. They have specific signatures in neutron scattering observables which have been well documented; see, e.g.\ Refs.~\onlinecite{Maleev62,Blume63,Ishida85,Roessli02,Maleyev06,Janoschek10}. Concerning another microscopic probe of magnetism, i.e.\ the $\mu$SR technique, recent works have evidenced that it is sensitive to unique characteristics associated with chiral magnetic structures \cite{Amato14,Dalmas16,Dalmas17} and excitations \cite{Yaouanc20}. Here we address the question of whether this technique is sensitive or not to paramagnetic chiral correlations. Based on the distinctive character of these correlations predicted by theory and observed in neutron scattering measurements \cite{Maleev62,Blume63,Roessli02}, and the nature of the coupling between the muon and the system, we find that they do not influence the zero-field $\mu$SR response when these correlations are dominated by modes around the Brillouin zone center.

Muons are spin 1/2 elementary particles. In the zero-field $\mu$SR experiments of interest here, muons fully polarized along the direction $Z$ are implanted into the specimen under study where they probe the dynamical microscopic magnetic fields arising from the magnetic moments of the system. The measured quantity is the evolution with time $t$ of the projection of their average polarization along the $Z$ axis: this is the so-called polarization function $P_Z(t)$. In paramagnets, $P_Z(t)$ typically decays according to an exponential function. The reader is referred to Ref.~\onlinecite{Yaouanc11} for further details about the $\mu$SR technique. We note that chiral magnets have been the object of many $\mu$SR studies both in their ordered and paramagnetic phases; see Refs.~\onlinecite{Hayano78,Kadono90,Braam15,Khasanov15,Lancaster16} for a few examples.

The organization of the paper is as follows.  Section~\ref{energy} introduces the minimal expression for the chiral magnetic free energy of a compound based on the DM interaction. In the next section (Sec.~\ref{Long}) we deal with the $\mu$SR zero-field polarization function in the paramagnetic state in general terms. The possible influence of chirality  on this function is investigated in Sec.~\ref{Chiral}. A summary and a discussion are given in Sec.~\ref{Summary_conclusion}. The text is supplemented with an appendix providing mathematical formulas.

\section{Chiral magnetic free energy}
\label{energy}

We consider a magnetic system in which the magnetic moment at the atomic sites varies so slowly that it can be considered as a continuous quantity. We introduce the magnetic free energy density averaged over the crystal volume $V$
\begin{eqnarray}
{\mathcal E} = \frac{1}{V} \int_V F({\bf r})\, {\rm d}^3{\bf r},
\label{energy_1}
\end{eqnarray}
where $F({\bf r})$ is the local energy density. Following Bak and Jensen \cite{Bak80}, the minimal chiral free energy is given by the sum of three terms:
\begin{eqnarray}
  F({\bf r}) & = & \frac{A} {2}  (S_x^2 + S_y^2 + S_z^2) + D {\bf S} \cdot (\nabla  \times {\bf S}) \cr
& + & \frac{B_1}{2} \left[(\nabla S_x)^2 +  (\nabla S_y)^2 +  (\nabla S_z)^2 \right].  
\label{energy_2}
\end{eqnarray}
We denote the spin density at position vector ${\bf r}$ as ${\bf S}({\bf r})$. The spin components are expressed in the reference frame $(x,y,z )$ of the crystal. In order of appearance on the right hand-side of Eq.~\ref{energy_2}, we have the Landau term, the antisymmetric Dzyaloshinski-Moriya (DM) exchange interaction, and the symmetric Heisenberg exchange interaction. This simplified model neglects other interactions which are possibly present, e.g.\ a weak exchange anisotropy or the crystal field energy. Their presence would not change our result. The Ginzburg-Landau framework we use is justified since only long wavelength modes matter in a compound with DM interaction.

Introducing the Fourier series
\begin{eqnarray}
{\bf S} ({\bf r}) = \sum_{\bf q} {\bf S}_{\bf q} \exp(i {\bf q} \cdot {\bf r}),
 \label{energy_4}
\end{eqnarray}
where
\begin{eqnarray}
{\bf S}_{\bf q} = \frac{1}{V} \int_V {\bf S}({\bf r}) \exp(-i {\bf q} \cdot {\bf r})\,{\rm d}^3{\bf r},
\label{energy_5}
\end{eqnarray}
Eq.~\ref{energy_1} writes 
\begin{eqnarray}
{\mathcal E} & = & \sum_{\bf q}
  \left[ \left (A + B_1 q^2  \right ) \frac{|{\bf S}_{\bf q}|^2}{2}  
    + i D {\bf q} \cdot ({{\bf S}}_{\bf q} \times {{\bf S}}_{-{\bf q}}) \right].
\label{energy_6}
\end{eqnarray}
For convenience we will consider ${\bf q}$ as a continuous variable from now on. Although conventional, Eq.~\ref{energy_6} is not suitable for our purpose. Since the DM interaction is antisymmetric, it is propitious to consider the operator
\begin{eqnarray}
P^{\alpha \beta}_{{\rm T}_2}({\bf q}) = \sum_\gamma \epsilon^{\alpha \gamma \beta} \frac{q^\gamma}{q},
\label{projector_T2}
\end{eqnarray}
where $\epsilon^{\alpha \gamma \beta}$ is the Levi-Civita symbol. Nicely enough, $P^{\alpha \beta}_{{\rm T}_2}({\bf q})$ depends only on the orientation of ${\bf q}$ and not its modulus. Recalling that ${\bf q} \cdot [{\bf S}({\bf q}) \times {\bf S}(-{\bf q})]$ = $- {\bf S}({\bf q})\cdot [{\bf q} \times {\bf S}(-{\bf q})]$ and with the help of Eq.~\ref{projectors}, we derive
\begin{eqnarray}
{\mathcal E} &  = & V \int {\bf S}^T({\bf q})
\left[ \frac{A + B_1 q^2  }{2} \boldsymbol{I}
  - i D q \boldsymbol{P}_{{\rm T}_2}({\bf q})  
\right]{\bf S}(-{\bf q})\,\frac{{\rm d}^3{\bf q}}{(2\pi)^3},\cr & &
\label{energy_10}
\end{eqnarray}
Here, $\boldsymbol{I}$ is the identity operator and ${\bf S}^T({\bf q})$ is the transpose of column vector ${\bf S}({\bf q})$. 

\section{Paramagnetic zero-field polarization function}
\label{Long}

In this section we describe the zero-field $\mu$SR polarization function in the paramagnetic state in general terms. We first establish the expressions of the spin-correlation functions needed for the computation of $P_Z(t)$ which follows. 

\subsection{Spin correlation function}
\label{Long_correlation}

The derivation of an expression for the $\mu$SR spin-lattice relaxation rate  (see Sec.~\ref{Long_rate}) requires the spin correlation tensor. In general terms, a component of this tensor is defined as
\begin{eqnarray}
\Lambda^{\alpha\beta}({\bf q},\omega) & = & \langle S^\alpha({\bf q},\omega) S^\beta(-{\bf q}) \rangle
\label{Long_correlation_1}
\end{eqnarray}
where $\alpha$ and $\beta$ denote Cartesian axes for the crystal reference frame \footnote{Here we consider the correlation tensor defined in Eq.~\ref{Long_correlation_1} rather than the usual symmetrized quantity $\Lambda^{\alpha\beta}({\bf q},\omega) = \left[ \protect\langle S^\alpha({\bf q},\omega) S^\beta(-{\bf q}) \protect \rangle  + \protect\langle S^\beta(-{\bf q}) S^\alpha({\bf q},\omega) \protect\rangle \right] /2$ \cite{Yaouanc11}. Since we restrict ourselves to the paramagnetic phase in zero-field, time reversal symmetry holds and the two terms of the symmetrized tensor are equal. This tensor is therefore equal to that defined in Eq.~\ref{Long_correlation_1}.}. We shall only need the tensor at angular frequency $\omega = 0$ since the paramagnetic state in zero field is considered. From the symmetry property of the free energy expressed in reciprocal space (Eq.~\ref{energy_10}) the following decomposition of the correlation function holds: 
\begin{eqnarray}
 \boldsymbol{\it \Lambda}({\bf q},\omega = 0) 
  & = & \Lambda^{\rm I}(q,\omega = 0)\boldsymbol{I} +  \Lambda^{{\rm T}_2}(q,\omega = 0)\boldsymbol{P}_{{\rm T}_2}({\bf q}).\cr & &
\label{Long_correlation_3}
\end{eqnarray}
As shown in Ref.~\onlinecite{Grigoriev05}, this is justified since we are interested by the small $q$ limit. The first term on the right hand side of Eq.~\ref{Long_correlation_3} is the isotropic part of the correlation which depends solely on the modulus of ${\bf q}$. The second term is the chiral contribution. Its dependence on the orientation of ${\bf q}$ is described by the operator $\boldsymbol{P}_{{\rm T}_2}({\bf q})$.

\subsection{Muon spin lattice relaxation function}
\label{Long_rate}

In the paramagnetic phase of a magnetic system, the zero-field relaxation function is an exponential function
\begin{eqnarray}
  P_Z(t) & = & \exp(-\lambda_Z t),
\label{Polarization}
\end{eqnarray}
characterized by the relaxation rate $\lambda_Z$ expressed as (Refs.~\onlinecite{Dalmas97,Yaouanc11})
\begin{eqnarray}
  \lambda_Z & = & \frac{{\mathcal D}}{2} 
  \int \sum_{\gamma,\gamma^\prime} \left[G^{X \gamma}({\bf q})G^{\gamma^\prime X}(-{\bf q})
    \right. \label{lambdaZ_single_1} \\ & &
    \hspace{10mm} + \left. G^{Y \gamma}({\bf q})G^{\gamma^\prime Y}(-{\bf q})\right]\Lambda^{\gamma \gamma^\prime}({\bf q},\omega =0) \frac{{\rm d}^3{\bf q}}{(2 \pi)^3}.
\nonumber
\end{eqnarray}
We have defined the constant ${\mathcal D} = \left ({\mu_0}/{4 \pi}  \right )^2 \gamma_\mu^2 g^2 \mu_{\rm B}^2/{v_{\rm c}}$,  where $g$ is the spectroscopic factor of the magnetic moments, $\mu_{\rm B}$ is the Bohr magneton, $v_{\rm c}$ is the unit cell volume, and $\gamma_\mu = 8.51616 \times 10^8$~rad\,s$^{-1}$\,T$^{-1}$ is the muon gyromagnetic ratio. The integral extends over the first Brillouin zone. Tensor $\boldsymbol{\it G}({\bf q})$ accounts for the coupling between the muon spin and the spins in the crystal \footnote{Equation \ref{lambdaZ_single_1} is valid for a crystal. For a powder sample an angular average is required. Such an average does not hamper the conclusion of the Paper.}. This coupling is of dipolar origin. In metallic systems the additional interaction between the muon spin and the electronic spin density at the muon site is described {\em via} the Fermi contact field, and its effect is included in $\boldsymbol{\it G}({\bf q})$.

Equation \ref{lambdaZ_single_1} is written in the laboratory reference frame $(X,Y,Z)$, where $Z$ is the direction along which the muon polarization is monitored (Sec.~\ref{Introduction}) and $X$ and $Y$ are two Cartesian directions perpendicular to each other and to $Z$.
Naturally, the components of the correlation tensor $\boldsymbol{\it \Lambda}$ are most conveniently expressed in the $(x,y,z )$ crystal frame. If the two frames do not coincide, rotations can be introduced in the expression of Eq.~\ref{lambdaZ_single_1}; see, e.g.\ Ref.~\onlinecite{Yaouanc11}.  Then all the components of the product of tensors $\boldsymbol{G}({\bf q})\boldsymbol{\it \Lambda}({\bf q},\omega = 0)\boldsymbol{G}(-{\bf q})$ in the crystal frame may be involved, rather than only the laboratory frame $XX$ and $YY$ components of this product as in Eq.~\ref{lambdaZ_single_1}.

As only long wavelength correlations matter in our modelling of the magnetic properties of the compound of interest, we are entitled to limit ourselves also to the long-wavelength limit of $G^{\alpha \beta}({\bf q})$. Referring to Refs.~\onlinecite{Yaouanc93,Yaouanc93a,Dalmas94},
\begin{eqnarray} 
G^{\alpha \beta}({\bf q}\to 0) & = & - 4 \pi \left (P^{\alpha \beta}_{\rm L} ({\bf q}) -C^{\alpha \beta} ({\bf q} = {\bf 0})
-\frac{r_\mu H}{4 \pi} \delta^{\alpha \beta}  
\right ),\cr & &
\label{lambdaZ_single_5}
\end{eqnarray}
where
\begin{equation}
  P^{\alpha \beta}_{\rm L} ({\bf q}) = \frac{q^\alpha q^\beta}{q^2},
\label{projector_L}
\end{equation}
is a component of the longitudinal projection operator $\boldsymbol{P}_{\rm L}({\bf q})$ and  $r_\mu H/4 \pi$ quantifies the effect of the Fermi contact field if present. The tensor $\boldsymbol{\it C}({\bf q} = {\bf 0})$ describes the analytical part at ${\bf q} = {\bf 0}$ of the dipole interaction between the muon and magnetic moments, while $\boldsymbol{P}_{\rm L}({\bf q})$ is only piecewise continuous at ${\bf q} = {\bf 0}$.

\section{Chiral contribution to the polarization function} 
\label{Chiral}

We now examine the contribution of the chiral correlations to the relaxation rate through the quantity
\begin{eqnarray}
 \int \boldsymbol{G}({\bf q})\Lambda^{{\rm T}_2}(q,\omega=0)\boldsymbol{P}_{{\rm T}_2}({\bf q})\boldsymbol{G}(-{\bf q}) \frac{{\rm d}^3{\bf q}}{(2\pi)^3}.
\label{integral}
\end{eqnarray}
The tensor $\boldsymbol{P}_{{\rm T}_2}({\bf q})$ together with the tensor $\boldsymbol{G}({\bf q})$ in the limit of small wavevectors of interest here (Eq.~\ref{lambdaZ_single_5}), depend only on the orientation of ${\bf q}$ and not its modulus whereas it is just the opposite for $\Lambda^{{\rm T}_2}(q,\omega=0)$. This remark suggests to compute the triple integral of Eq.~\ref{integral} in spherical coordinates. We start with the integrals over the polar and azimuthal angles. Given that $\boldsymbol{G}({\bf q})$ is the sum of $\boldsymbol{P}_{\rm L}({\bf q})$ and a constant, the expansion of the double product $\boldsymbol{G}({\bf q})\boldsymbol{P}_{{\rm T}_2}({\bf q})\boldsymbol{G}(-{\bf q})$ is a weighted sum of the components of four terms: $\boldsymbol{P}_{{\rm T}_2}({\bf q})$, $\boldsymbol{P}_{\rm L}({\bf q})\boldsymbol{P}_{{\rm T}_2}({\bf q})$, $\boldsymbol{P}_{{\rm T}_2}({\bf q})\boldsymbol{P}_{\rm L}({\bf q})$, and $\boldsymbol{P}_{\rm L}({\bf q})\boldsymbol{P}_{{\rm T}_2}({\bf q})\boldsymbol{P}_{\rm L}({\bf q})$. The last three terms vanish because of the orthogonality relation mentioned in Eq.~\ref{orthogonality}. Concerning the first term, its contribution to Eq.~\ref{integral} cancels since
\begin{eqnarray}
\int_0^{2\pi} \int_0^{\pi}P^{\gamma \gamma^\prime}_{{\rm T}_2}({\bf q}) \sin\theta\,{\rm d} \theta\,{\rm d} \phi  = 0,
\label{Chiral_5}
\end{eqnarray}
for all Cartesian components $\gamma$ and $\gamma^\prime$. 

We have therefore substantiated that the angular integrals of Eq.~\ref{integral} vanish. Finally we check that the integral over $q$ is finite. Recasting to the fluctuation-dissipation theorem in the limit $\hbar\omega/k_{\rm B}T \to 0$ (see, e.g.\ Ref.~\onlinecite{Yaouanc11}), we get
\begin{eqnarray}
\Lambda^{{\rm T}_2}(q,\omega=0) & \propto & \frac{\chi^{{\rm T}_2}(q)}{\Gamma^{{\rm T}_2}(q)},
\label{fluctuation_dissipation}
\end{eqnarray}
where $\chi^{{\rm T}_2}(q)$ is the chiral static susceptibility and $\Gamma^{{\rm T}_2}(q)$ is the associated linewidth. We expect the latter quantity to be independent of $q$ in the long wavelength limit above the critical temperature since the DM interaction violates the total spin conservation law \cite{Maleyev06}. Then from the $q$ dependence of $\chi^{{\rm T}_2}(q)$, see, e.g.\ Ref.~\onlinecite{Grigoriev05}, we find the radial integral to be finite.

In conclusion the quantity defined in Eq.~\ref{integral} is zero. We have therefore established that the chiral correlations do not contribute to the $\mu$SR relaxation rate.

\section{Summary and Discussion}
\label{Summary_conclusion}

We examined the influence of the chiral correlations on the zero-field $\mu$SR spectra measured in the paramagnetic phase of chiral magnets. The derivation is analytical and relies on the specific ${\bf q}$ dependence of chiral correlations. It is performed in the limit of small wavevectors at which these spin correlations dominate. This approximation is especially justified for magnets that order with a small propagation wavevector, such as the systems for which the chirality stems from the DM interaction. In addition, we note that when the temperature is high enough that the small ${\bf q}$ approximation might no longer be relevant, the chiral correlations tend to be suppressed \cite{Pappas09,Grigoriev10a,Pappas11}. Owing to the specific symmetry properties of the dipolar --- and possible Fermi contact field --- coupling between the muon spin and the chiral correlations, it is found that these correlations do not couple to the muon spin. As a consequence the muon-spin relaxation rate solely probes non-chiral correlations.

At this stage, it is instructive to review how the analytical derivation presented in Sec.~\ref{Chiral} can alternatively be inferred from an inspection of Eq.~\ref{lambdaZ_single_1}. As already mentioned, $\lambda_Z$ involves an integral over the polar and azimuthal angles of the components $XX$ and $YY$ of the product of tensors ${\boldsymbol G}({\bf q}) {\boldsymbol P}_{{\rm T}_2}({\bf q}){\boldsymbol G}(-{\bf q})$. For the former component as an example, this product can be viewed as the scalar product of vectors $G^{X\gamma}({\bf q})$ and ${\boldsymbol P}_{{\rm T}_2}({\bf q}) G^{\gamma^\prime X}(-{\bf q})$. Since ${\boldsymbol P}_{{\rm T}_2}({\bf q}) G^{\gamma^\prime X}(-{\bf q})$ is perpendicular to $G^{\gamma^\prime X}(-{\bf q})$ = $G^{\gamma^\prime X}({\bf q})$ = $G^{X\gamma^\prime}({\bf q})$ (see Eqs.~\ref{projectors} and \ref{lambdaZ_single_5} and Ref.~\footnote{As a general property, the ${\boldsymbol G}({\bf q})$ tensor is symmetric; see Ref.~\cite{Yaouanc11}.}) the scalar product vanishes as $\lambda_Z$ does.

Our result can be applied to the spin-lattice relaxation time $T_1$ measured in nuclear quadrupole resonance (NQR) experiments. As for $\mu$SR, this technique is not sensitive to chiral correlations \footnote{Strictly speaking, NQR does not probe the correlation functions at an angular frequency $\omega$ = 0 as $\mu$SR but at a value corresponding to the quadrupole splitting. However this splitting is extremely small compared to the relevant energies in $F({\bf r})$ or to the thermal energy.}. We are not aware of any NQR study of $1/T_1$ in the paramagnetic phase of a chiral magnet.

The conclusion about the absence of sensitivity of $\mu$SR to chiral correlations does not mean that this technique provides no information about other aspects of chiral magnetism. As far as the model helimagnet MnSi is concerned, the chirality of the zero-field magnetic order can be experimentally verified, provided the handedness of the crystal structure is known \cite{Amato14,Dalmas16}. The unique wavevector anisotropy of the dispersion relation of the helimagnon excitations in the ordered phase can also be probed \cite{Yaouanc20}.

\section*{Acknowledgments}

We thank C. Pappas for a discussion and B. Roessli and A. Maisuradze for a critical reading of the manuscript.

\appendix
\section{The $\boldsymbol{P}_{\rm L}({\bf q})$ and $\boldsymbol{P}_{{\rm T}_2}({\bf q})$ operators}
\label{sec:projectors}
We provide an insight into the operators respectively defined in Eqs.~\ref{projector_L} and \ref{projector_T2} and derive an orthogonality property which is a key for the material presented in Sec. \ref{Chiral}. 

First we note that the two operators are defined for any non-zero ${\bf q}$ vector. When applied to a ${\bf V}$ vector, the result $\boldsymbol{P}_{\rm L}({\bf q}){\bf V}$ is the vector collinear to $\hat{\bf q}\equiv {\bf q}/q$ with a length equal to the scalar product $\hat{\bf q}\cdot{\bf V}$. This definition justifies the name longitudinal projection operator given to $\boldsymbol{P}_{\rm L}({\bf q})$. Concerning $\boldsymbol{P}_{{\rm T}_2}({\bf q})$ this is the vectorial product $\hat{\bf q}\times {\bf V}$. The two relations summarize as
\begin{eqnarray}
  \boldsymbol{P}_{\rm L}({\bf q}){\bf V} & = & (\hat{\bf q}\cdot{\bf V})\hat{\bf q},\cr
 \boldsymbol{P}_{{\rm T}_2}({\bf q}) {\bf V} & = & \hat{\bf q}\times {\bf V}.
  \label{projectors}
\end{eqnarray}

Inspecting Eq.~\ref{projectors}, we notice that $\boldsymbol{P}_{{\rm T}_2}({\bf q}) {\bf V}$ is essentially a vector perpendicular to $\hat{\bf q}$ and ${\bf V}$ and that $\boldsymbol{P}_{\rm L}({\bf q})$ projects along the $\hat{\bf q}$ vector. Therefore,
\begin{eqnarray}
  \boldsymbol{P}_{\rm L}({\bf q})\boldsymbol{P}_{{\rm T}_2}({\bf q}) & = &
  \boldsymbol{P}_{{\rm T}_2}({\bf q})\boldsymbol{P}_{\rm L}({\bf q}) = 0.
\label{orthogonality}
\end{eqnarray}
Obviously, an alternative derivation of these two relations can be algebraically inferred from the definition of the two operators (Eqs.~\ref{projector_L} and \ref{projector_T2}).

\bibliography{reference}
\end{document}